\documentclass[conference]{IEEEtran}

\usepackage{cite}
\usepackage{graphicx}
\usepackage{epstopdf} 
\usepackage{multirow}
\usepackage{pbox}
\graphicspath{{./figures/}}
\usepackage{xcolor}
\usepackage[official]{eurosym}
\usepackage[nolist]{acronym}

\title{LTE Standard-Compliant D2D Communication: Software-defined Radio Implementation and Evaluation}

\author{Shekhar Jain, Yi Zhang and Luiz A. DaSilva\\  
Trinity College Dublin, Ireland\\
email:
\{sjain, zhangy8, dasilval\}@tcd.ie
}

\begin{document}

\maketitle
\begin{abstract}
In this paper, we describe the design and implementation of \ac{D2D} communication functionality for \ac{LTE}. To our knowledge, this is the first such implementation that is compliant with the \ac{LTE} Release 12 standard.
We implement our system on a software-defined radio \ac{SDR} testbed, augmenting the open-source LTE eNodeB and \ac{UE} implementation provided by the srsLTE software suite. Our measurements demonstrate the cell extension capabilities of \ac{D2D} and quantify the \ac{SNR} and throughput obtained by a subscriber when directly served by the eNodeB and when provided connectivity through the relay \ac{UE}. Our implementation of sidelink, relaying, and mode selection functionality enables experimentation and prototyping of \ac{D2D} communication that can assist in standardization, research, and development.

\end{abstract}



\section{Introduction}
\label{sec:introduction}

4G mobile network standards are being extended to support \ac{D2D} communications, functionality that is also expected to be present in 5G~\cite{RSwhitepaper}. Standardization efforts from the \ac{3GPP} are ongoing to provide \ac{LTE} with \ac{D2D} and relaying capabilities, aiming to achieve  coverage extension and improved transmission reliability ~\cite{3GPP}. D2D communication can also support proximity services and public safety use of LTE networks. 

\begin{figure}[!t]
	\centering
	\includegraphics[width=0.4\textwidth, trim = 0mm 0mm 0mm 0mm, clip ]{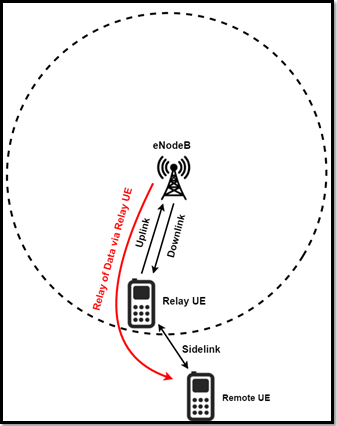}  
	\caption{Sidelink communication in LTE networks.}
	\label{fig:car_network}
\end{figure}

The \ac{PHY}  communication link between two UEs is referred to as a \emph{sidelink}, standardized in LTE Release 12~\cite{3GPP}, involving user discovery, synchronization, and data transmission. In some locations (e.g. in the edge of cells) where coverage from the \ac{LTE} base station cannot be guaranteed because of low \ac{SNR}, it is possible to use a UE as a relay from the eNodeB to a more remote UE, as illustrated in Fig.~\ref{fig:car_network}. As the user may exhibit mobility and channel conditions change from time to time, dynamic mode selection functionality between infrastructure mode (eNodeB-to-UE) and \ac{D2D} mode (UE-to-UE) is needed.

In this paper we develop the first \ac{SDR} PHY layer sidelink  implementation compliant with the \ac{3GPP} \ac{LTE} standard (Release 12). We also implement functionality to relay data received from an eNodeB (in the downlink) to the remote \ac{UE} (through the sidelink), as well as mode selection functionality to allow the remote UE to switch between sidelink mode and downlink mode according to network conditions.  We extend an open source SDR platform, srsLTE~\cite{srsLTE_link}, with our implementation of the sidelink.
This allows us to dynamically control the radio parameters, e.g. frequency, gain, and \ac{MCS} index. 
We also develop controllers with Graphical User Interfaces (GUIs) that allow experimenters to 
change radio parameter settings on the fly. Our testbed employs the \ac{USRP} platform~\cite{Ettus}. We report measurement results of SNR, throughput and Block Error Rate (BLER) and discuss how these can be used as the criteria for mode selection between direct downlink and relaying through the sidelink. Our implementation of the sidelink, relaying, and mode selection in open source \ac{LTE}  software enables experimentation and prototyping that can assist in standardization, research and development.

D2D communication has been a topic of much recent research, as discussed in the survey~\cite{D2D_survey}. However, only a few research works have developed testbed implementations for testing the performance of systems that rely on \ac{D2D} communications. For instance,
the authors of~\cite{SiChBo2016} present the measurement results of power consumption and \ac{RSSI} for their \ac{D2D} communication in a testbed of sensor nodes following the IEEE 802.15.4 standard. The authors of~\cite{ZhZhZh2016} integrate \ac{D2D} communication and \ac{SDN} by adding \ac{MAC} layer functionality (TDMA, SCMA, etc.) to the basic \ac{OFDM}-based \ac{PHY} layer transmission; the \ac{SDN} controller is then implemented on top of the \ac{MAC} layer protocol. The authors of~\cite{ChLiLi2016} investigate cross-link interference management in a proposed multi-user \ac{D2D} network scenario. The experiment in the paper was carried out by using basic OFDM physical transmission.
However, none of the works mentioned above has produced sidelink implementations that conform to the current 3GPP LTE standardization of D2D communication. To the best of our knowledge, ours is the first PHY layer implementation of D2D communication that is compliant with LTE Release 12 specifications.

For implementing the sidelink communication and relay functionality, and to make it compliant with the LTE standard, we have used the srsLTE library~\cite{srsLTE_link}, an open source \ac{PHY} layer platform for software-defined radio implementation of the \ac{FDD} mode of \ac{LTE}~\cite{Gomez-Miguelez2016}. The library follows a modular approach, which helps in combining different PHY layer \ac{DSP} components together without the need to change every element. We have extended the code, implementing the sidelink and relay functionalities while also making use of existing components in srsLTE, namely the uplink and downlink modules, which are compliant to 3GPP LTE Release 8.

This work is conducted in the context of the Horizon 2020 (H2020) \ac{eWINE} project, which focuses on experimental research towards flexible, on-demand end-to-end wireless connectivity. This elastic network connectivity requires the dynamic reallocation of network resources. Our work in this paper implements D2D capabilities in the LTE network which can then be used to dynamically establish a sidelink or a link directly with the eNodeB, depending on current network conditions. 

\section{Design and Implementation}
\label{sec:Implementation}

As described in the previous section, our design and implementation focuses on the LTE sidelink and relay functionality. In this section, we outline the major technical challenges in implementing this functionality. The first challenge is to develop the sidelink PHY layer data transmission channels, which adopt a \ac{SC-FDMA} waveform, according to the \ac{LTE} standard. The second challenge is to build the relay functionality for the UE. The third challenge is to design the mode selection functionality in the remote UE to establish either a sidelink (with the relay UE) or a downlink connection (with the eNodeB).

\subsection{Implementation of \ac{PSSCH} with \ac{SC-FDMA}}

\begin{figure}[!t]
	\centering
	\includegraphics[width=0.5\textwidth, trim = 0mm 0mm 0mm 0mm, clip ]{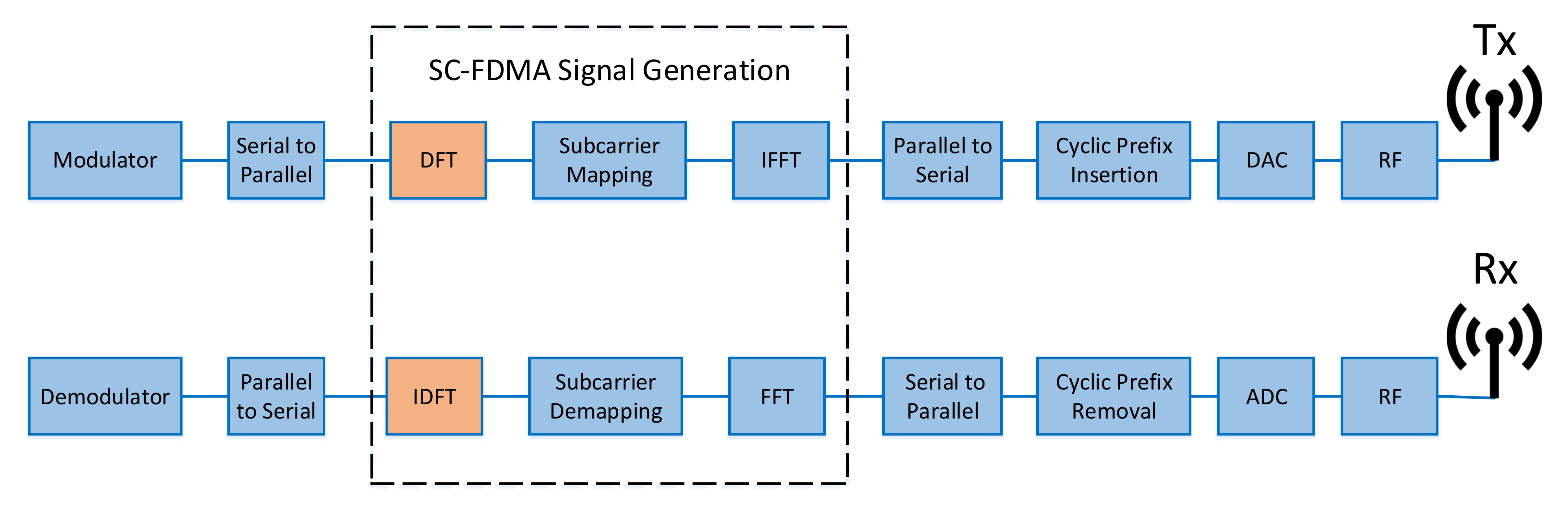}  
	\caption{Block diagram of SC-FDMA transmitter (Tx) and receiver (Rx).}
	\label{fig:sc_fdma}
\end{figure}

The \ac{PSSCH} is the \ac{PHY} channel for data transmission on the sidelink. 
According to the \ac{3GPP} \ac{LTE} standard Release 12, the PSSCH must use \ac{SC-FDMA} for data transmission. The main  difference in implementation between \ac{SC-FDMA} and \ac{OFDM} (which is used in the downlink) is that an additional \ac{DFT} is needed in the transmitter and an additional \ac{IDFT} is needed in the receiver, as shown in Fig.~\ref{fig:sc_fdma}. We implement the \ac{PSSCH} by inserting the \ac{DFT} and \ac{IDFT} modules, available at the FFTW3 and srsLTE software library, into the \ac{OFDM} radio \ac{DSP} chain for the \ac{PSSCH}. 

A \ac{PSCCH} is also needed for accomplishing the functionality of \ac{PSSCH}. In our sidelink implementation, three symbols per subframe are allocated for the \ac{PSCCH}, and \ac{SCI} is included in these symbols.
The \ac{SCI} may include information on the \ac{MCS}, a group destination identifier, the resource block assignment and hopping resource allocation, a frequency hopping flag, etc. In our implementation we include minimum functionality for the \ac{PSCCH}, supporting flexible \ac{MCS} selection by the sidelink controller (to be described in the following sections). For the moment, we 
assign the entire set of resource blocks in the frequency channel of the sidelink to a single pair of \acp{UE}. 

\subsection{Implementation of relay functionality in the UE}

\begin{figure}[!t]
	\centering
	\includegraphics[width=0.5\textwidth, trim = 0mm 0mm 0mm 0mm, clip ]{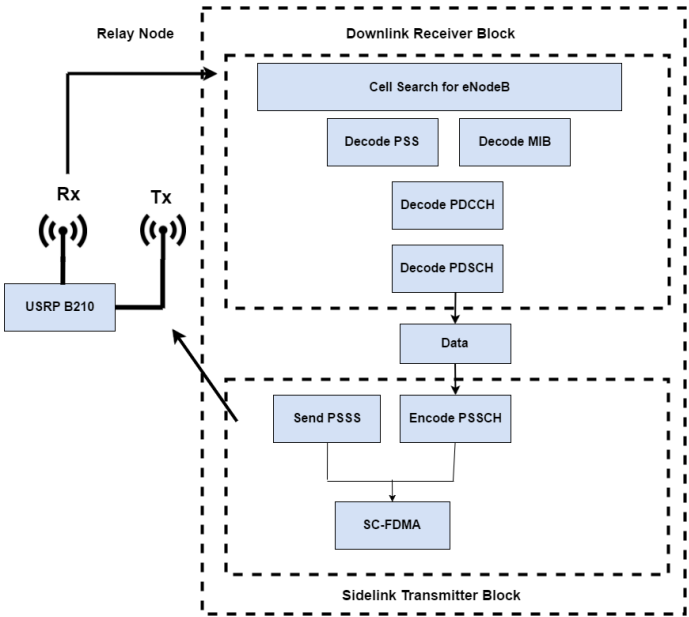}  
	\caption{Block diagram of the relay UE.}
	\label{fig:block_relay_UE}
\end{figure}

The relay \ac{UE} is called upon to relay the data received in the downlink to the remote UE, communicating over the sidelink. In order to run the downlink and sidelink in parallel, we create two different threads, which individually perform the two sets of operations without interrupting each other. The internal functionality of the relay \ac{UE} is shown in Fig.~\ref{fig:block_relay_UE}. A \ac{UE} must  establish the link with the eNodeB, first searching for beaconing signals periodically transmitted by the base station (in the figure, this process is referred to as a cell search). After the downlink is established, the \ac{UE} must decode the \ac{PSS} and \ac{MIB} signals to synchronize with the eNodeB. Then, it decodes the \ac{PDCCH} and the \ac{PDSCH} signals to receive data, which subsequently needs to be forwarded to the remote \ac{UE} over the sidelink. This data is then transmitted over the \ac{PSSCH} using \ac{SC-FDMA}. The \ac{PSSS} signal is also generated simultaneously, for synchronization between  two UEs that have established a sidelink. 

\subsection{Implementation of mode switching}

We implement functionality for the remote \ac{UE} that enables it to switch between sidelink and downlink modes of operation in real time. A controller is needed to perform this mode switching with minimum latency. We implement our controller with two separate threads: one calls up the \ac{OFDM} receiver for the downlink and the other one calls up the receiver for the sidelink. The threads run the individual Linux bash commands which can boot either the downlink or sidelink program automatically. For experimental purposes, we build a \ac{GUI} on top of this controller that can be used by an experimenter to manually switch between operation modes, as further described in  section~\ref{sec:Additional_Contribution}.

\section{Experimental Results}
\label{sec:Results}

\begin{figure}[!t]
	\centering
	\includegraphics[width=0.48\textwidth, trim = 0mm 0mm 0mm 0mm, clip ]{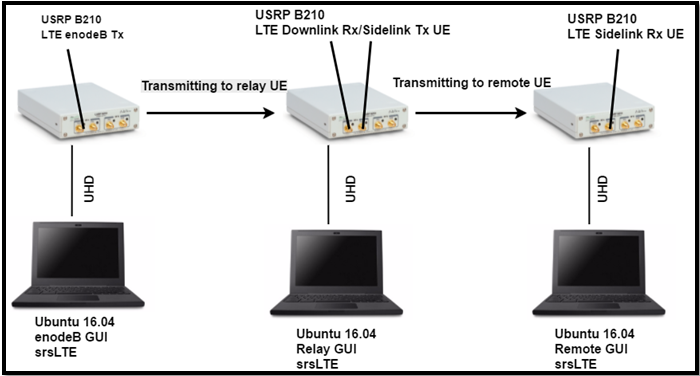}  
	\caption{Experimental set-up.}
    	\label{fig:testbed_set_up}
\end{figure}

To verify the correct operation of the solutions described above, and to collect performance measurement results, we have set up an experiment as shown in Fig.~\ref{fig:testbed_set_up}. The figure shows the eNodeB on the left hand side, a \ac{UE} that serves as a relay in the middle, and the remote UE on the right hand side. For the experiments reported here, we adopted the \ac{USRP} reconfigurable radio model B210 and the \ac{UHD} interface. The \ac{USRP} B210 hardware supports a 30.72 MHz clock, matching \ac{LTE} sampling frequencies and enabling the decoding of signals from live \ac{LTE} base stations. Our experiment consists of three \ac{USRP} B210 and three laptops running srsLTE software, extended with our sidelink and relay implementation. Each \ac{USRP} connects to a different laptop through a \ac{UHD} driver and USB 3.0. 

\begin{figure}[!t]
	\centering
	\includegraphics[width=0.48\textwidth, trim = 0mm 0mm 0mm 0mm, clip ]{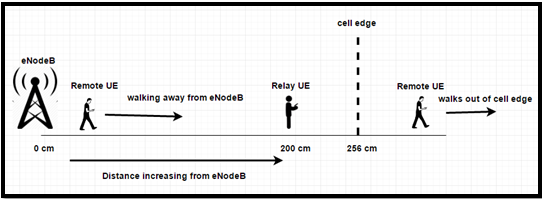}  
	\caption{Measurement scenario.}
    	\label{fig:Results_Scenario}
\end{figure}

\begin{figure}[!t]
	\centering
	\includegraphics[width=0.48\textwidth, trim = 0mm 0mm 0mm 0mm, clip ]{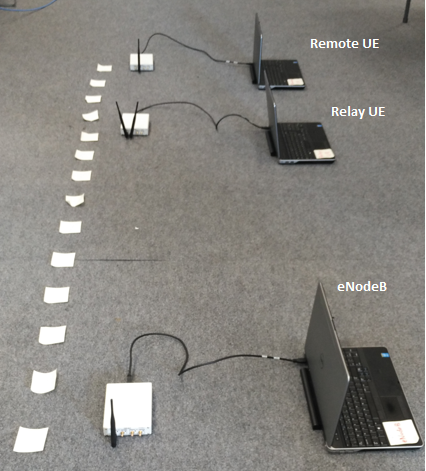}  
	\caption{Measurement Setup.}
    	\label{fig:measurement_setup}
\end{figure}

In our measurements we maintain the locations of eNodeB and relay \ac{UE} fixed and move the remote UE, starting from the position of the eNodeB, towards the edge of the cell
as depicted in Fig.~\ref{fig:Results_Scenario}.
A photograph of the measurement set-up is shown in Fig.~\ref{fig:measurement_setup}.
The measurements focus on how the \ac{SNR} and throughput vary with the position of the remote \ac{UE}. In the tests, a downlink signal is transmitted from the eNodeB to the remote \ac{UE} and sidelink signal is transmitted from the relay \ac{UE} to the remote \ac{UE}. To accommodate the range restrictions of the RF hardware and the space available in the laboratory where the experiments took place, we fixed the transmitter gain for the eNodeB at 55 dB, yielding a cell boundary at approximately 256 cm.

The following subsections describe the measurement results.

\subsection{SNR measurements and mode selection}

\begin{figure}[!t]
	\centering
	\includegraphics[width=0.48\textwidth, trim = 0mm 0mm 0mm 0mm, clip ]{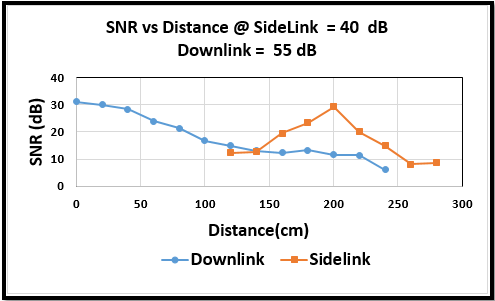}  
	\caption{SNR vs. distance for 40 dB sidelink transmitter gain.}
    	\label{fig:SNR_vs_Distance_40}
\end{figure}

Fig.~\ref{fig:SNR_vs_Distance_40} shows the average \ac{SNR} as a function of distance, measured from the location of the eNodeB to the remote UE, for both the sidelink (orange curve) and downlink (blue curve), when the transmitter gain of the sidelink is fixed at 40 dB. (This is lower than the transmitter gain for the downlink because the relay \ac{UE} is supposed to have lower power consumption for transmission and lower coverage than an eNodeB.) 
It is clear that, up to around 160 cm from the eNodeB, the downlink offers superior SNR to the sidelink; past that point, establishing a sidelink to a relay UE provides an SNR advantage. Beyond a distance of around 240 cm from the eNodeB, the only option available to the remote UE is to set up a sidelink: this illustrates the range extension capabilities of D2D.

\begin{figure}[!t]
	\centering
	\includegraphics[width=0.48\textwidth, trim = 0mm 0mm 0mm 0mm, clip ]{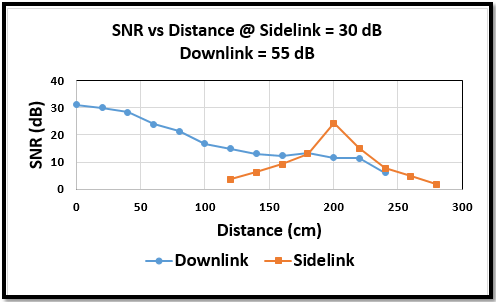}  
	\caption{SNR vs. distance for 30 dB sidelink transmitter gain.}
    	\label{fig:SNR_vs_Distance_30}
\end{figure}

Fig.~\ref{fig:SNR_vs_Distance_30} shows  similar results of average \ac{SNR} versus distance, but in this case the sidelink transmitter gain has been set to 30 dB. We can see that around 180 cm is the switching point from downlink to sidelink, if the objective is to maximize SNR. We can also see that the \ac{SNR} for the sidelink is lower than that reported in Fig.~\ref{fig:SNR_vs_Distance_40}, and the effective range extension is also reduced, reflecting the lower setting for the sidelink gain.

Each point in Fig.~\ref{fig:SNR_vs_Distance_40} and Fig.~\ref{fig:SNR_vs_Distance_30} is the average of 1000 collected measurements. 
We also report the standard deviation, the width of the 95\% confidence internal, minimum, and maximum SNR in the Tables \ref{table:1}, \ref{table:2}, and \ref{table:3}. 
Statistical analysis reveals that, for each value of distance, the \ac{SNR} varies very little and hence we have very tight confidence intervals.

\begin{table}[ht]
\caption{Statistical analysis of the SNR for the sidelink, at 30 dB transmitter gain.}
\begin{tabular}{|c|c|c|c|c|c|}
\hline
 \textbf{Distance} & \textbf{Average} & \textbf{Standard} & \textbf{Confidence} & \textbf{Max} & \textbf{Min} \\ 
  \textbf{(cm)} &  \textbf{(dB)} & \textbf{Deviation (dB)} & \textbf{Interval (dB)} & \textbf{(dB)} & \textbf{(dB)} \\
 \hline
 120 & 3.6403  & 0.0812 & 0.0050  & 3.9  & 3.4 \\  
 140 & 6.2020   & 0.1803 & 0.0112 & 6.7  & 5.1 \\ 
 160 & 9.1434  & 0.2460  & 0.0152 & 9.7  & 8.4 \\ 
 180 & 12.9839 & 0.1278 & 0.0079 & 13.3 & 12.1 \\ 
 200 & 24.2176 & 0.3149 & 0.0195 & 24.9 & 23.2 \\ 
 220 & 15.087  & 0.2635 & 0.0163 & 15.9 & 14.1 \\ 
 240 & 7.6485  & 0.2282 & 0.0141 & 8.2  & 6.8 \\ 
 260 & 4.7130   & 0.2316 & 0.0144 & 5.1  & 2.7 \\ 
 280 & 1.9099  & 0.5534 & 0.0343 & 2.2  & 1.5 \\ 
 \hline
\end{tabular}
\label{table:1}
\end{table}

\begin{table}[ht]
\caption{Statistical analysis of the SNR for the sidelink, at 40 dB transmitter gain.}
\begin{tabular}{|c|c|c|c|c|c|}
\hline
 \textbf{Distance} & \textbf{Average} & \textbf{Standard} & \textbf{Confidence} & \textbf{Max} & \textbf{Min} \\ 
  \textbf{(cm)} &  \textbf{(dB)} & \textbf{Deviation (dB)} & \textbf{Interval (dB)} & \textbf{(dB)} & \textbf{(dB)} \\ 
 \hline
 120 & 12.3448 & 0.3534 & 0.0219 & 13.5 & 11.5 \\  
 140 & 12.6533 & 0.1633 & 0.0101 & 13.2 & 12.0 \\ 
 160 & 19.5281 & 0.2237 & 0.0139 & 20.1 & 18.8 \\ 
 180 & 23.6827 & 0.4287 & 0.0266 & 24.6 & 22.6 \\ 
 200 & 29.2872 & 0.8827 & 0.0547 & 30.7 & 26.8\\ 
 220 & 19.9731 & 0.2355 & 0.0146 & 20.7 & 19.2 \\ 
 240 & 14.9266 & 0.1093 & 0.0068 & 15.2 & 14.6 \\ 
 260 & 8.1482  & 0.1636 & 0.0101 & 8.5  & 7.3 \\ 
 280 & 8.5819  & 0.2286 & 0.0142 & 9.1  & 7.3 \\ 
 \hline
\end{tabular}
\label{table:2}
\end{table}

\begin{table}[ht]
\caption{Statistical analysis of the SNR for the downlink, at 55 dB transmitter gain.}
\begin{tabular}{|c|c|c|c|c|c|}
\hline
 \textbf{Distance} & \textbf{Average} & \textbf{Standard} & \textbf{Confidence} & \textbf{Max} & \textbf{Min} \\ 
  \textbf{(cm)} &  \textbf{(dB)} & \textbf{Deviation (dB)} & \textbf{Interval (dB)} & \textbf{(dB)} & \textbf{(dB)} \\
 \hline
 0   & 31.1044 & 1.5805 & 0.0980  & 34.0   & 27.9 \\  
 20  & 29.8855 & 1.1352 & 0.0704 & 32.2 & 27.2 \\ 
 40  & 28.3759 & 0.7280  & 0.0451 & 29.9 & 26.8 \\ 
 60  & 23.8751 & 1.0637 & 0.0659 & 25.3 & 14.7 \\ 
 80  & 21.3312 & 1.0483 & 0.0650  & 23.1 & 16.4\\ 
 100 & 16.6697 & 1.5552 & 0.0964 & 18.3 & 12.4 \\ 
 120 & 14.8763 & 0.7983 & 0.0495 & 17.3 & 12.6 \\ 
 140 & 13.1189 s& 0.9459 & 0.0586 & 15.5 & 9.0\\ 
 160 & 12.2334 & 0.8596 & 0.0533 & 15.1 & 9.2 \\ 
 180 & 13.2411 & 0.7260  & 0.0450  & 15.2 & 11.0 \\ 
 200 & 11.6577 & 0.8913 & 0.0552 & 13.5 & 9.1 \\ 
 220 & 11.3320  & 0.3551 & 0.0220  & 12.3 & 10.2 \\ 
 240 & 6.0951  & 0.3073 & 0.0190  & 6.9  & 3.2\\
 
 \hline
\end{tabular}
\label{table:3}
\end{table}

\subsection{Throughput measurements and mode selection}

\begin{figure}[!t]
	\centering
	\includegraphics[width=0.48\textwidth, trim = 0mm 0mm 0mm 0mm, clip ]{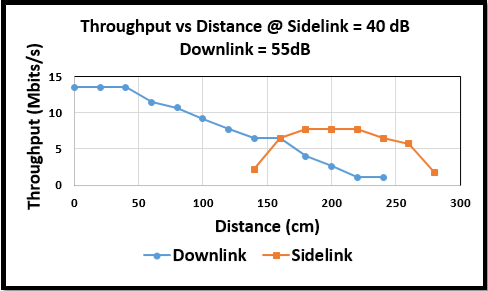}  
	\caption{Maximum throughput versus distance, for 40 dB sidelink gain.}
    	\label{fig:Throughput_vs_Distance_40}
\end{figure}

\begin{figure}[!t]
	\centering
	\includegraphics[width=0.48\textwidth, trim = 0mm 0mm 0mm 0mm, clip ]{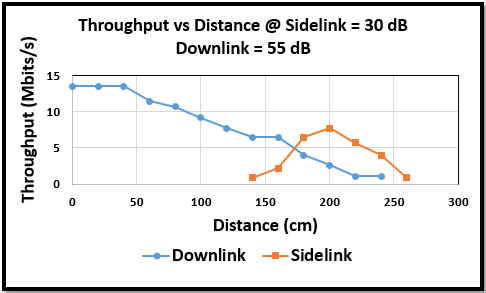}  
	\caption{Maximum throughput versus distance for 30 dB sidelink gain.}
    	\label{fig:Throughput_vs_Distance_30}
\end{figure}

We also measure the maximum throughout using the same experimental setup shown in Fig.~\ref{fig:Results_Scenario} and Fig.~\ref{fig:measurement_setup}. 
In the \ac{3GPP} \ac{LTE} standard, the throughput is determined by the \ac{MCS} index and number of \ac{PRBs} utilized for transmission. We allocate 25 \ac{PRBs} for both downlink and sidelink in our experiment. Therefore, at each measurement location we seek to reach the maximum throughput by increasing the \ac{MCS} index of the transmitters, for the downlink and sidelink, and record the maximum throughput while the observed \ac{BLER} remains at zero. The highest modulation for the downlink is 64 \ac{QAM} and the highest modulation for the sidelink is 16 \ac{QAM} according to the \ac{LTE} standard. Figs.~\ref{fig:Throughput_vs_Distance_40} and~\ref{fig:Throughput_vs_Distance_30} show how the throughput varies according to the position of the remote \ac{UE}; the trends follow closely those reported for the \ac{SNR}.



\section{Controller GUI for dynamic parameter settings}
\label{sec:Additional_Contribution}

We also built a user interface that allows experimenters to change the radio parameters, such as frequency, gain, and \ac{MCS}, on the fly for the eNodeB, relay \ac{UE} and remote \ac{UE}. We used the \textit{pThread} library in C to create threads that receive external commands from users and translate them into parameters settings in the srsLTE implementation. We used the GTK+ 3.0 library in C to implement the GUIs, and we implemented the communication between the \ac{GUI} and our \ac{LTE} software (based on a client-server scheme) using Linux socket interprocess communication libraries. The elements of our implementation include: 


\subsubsection{Controller GUI for eNodeB}
\begin{figure}[!t]
	\centering
	\includegraphics[width=0.45\textwidth, trim = 0mm 0mm 0mm 0mm, clip ]{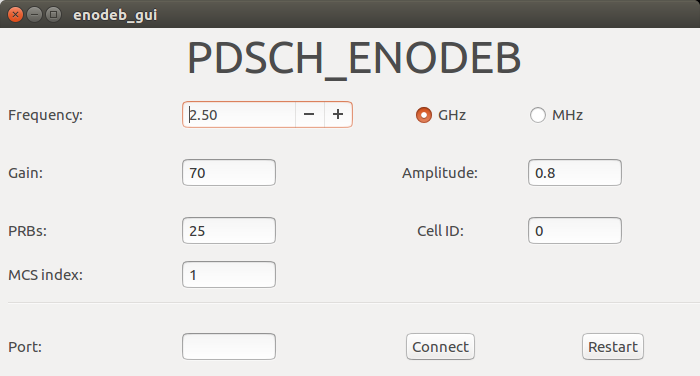}  
	\caption{Controller GUI for eNodeB.}
	\label{fig:enodeb_GUI}
\end{figure}

This eNodeB controller \ac{GUI} (shown in Fig.~\ref{fig:enodeb_GUI}) has the functionality to change frequency, gain, \ac{PRBs}, amplitude, \ac{MCS} index and cell ID.  

\subsubsection{Controller GUI for relay UE}

\begin{figure}[!t]
	\centering
	\includegraphics[width=0.45\textwidth, trim = 0mm 0mm 0mm 0mm, clip ]{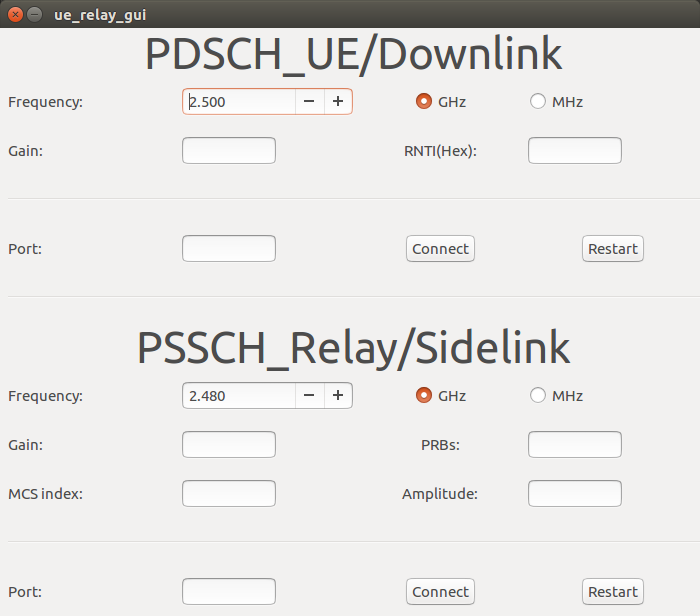}  
	\caption{Controller GUI for the relay UE, showing the control of both downlink and sidelink parameters.}
	\label{fig:relay_GUI}
\end{figure}

The controller \ac{GUI} for the relay \ac{UE}, shown in Fig.~\ref{fig:relay_GUI}, provides an interface for the downlink receiver and the sidelink transmitter, as the relay UE performs both functions at the same time. For the downlink receiver, the experimenter can change frequency, gain, and \ac{RNTI}. For the sidelink transmitter, the experimenter can change frequency, gain, amplitude, \ac{PRBs} and \ac{MCS} Index. This GUI has two different socket port options to connect to the \ac{LTE} software, so that the downlink and sidelink parameters are controlled separately via different Linux socket ports, providing additional flexibility to the experimenter. 

\subsubsection{Controller GUI for remote UE}

\begin{figure}[!t]
	\centering
	\includegraphics[width=0.45\textwidth, trim = 0mm 0mm 0mm 0mm, clip ]{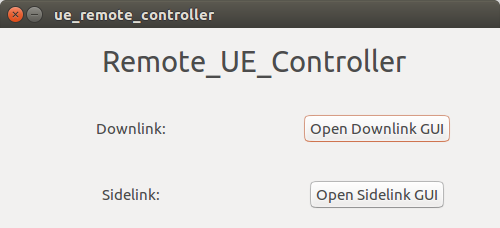}  
	\caption{Controller GUI for the remote UE: mode selection.}
	\label{fig:remote_controller}
\end{figure}

\begin{figure}[!t]
	\centering
	\includegraphics[width=0.45\textwidth, trim = 0mm 0mm 0mm 0mm, clip ]{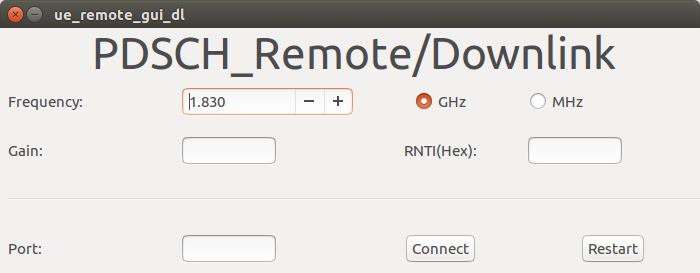}  
	\caption{Controller GUI for the remote UE using the downlink.}
	\label{fig:remote_downlink_GUI}
\end{figure}

\begin{figure}[!t]
	\centering
	\includegraphics[width=0.45\textwidth, trim = 0mm 0mm 0mm 0mm, clip ]{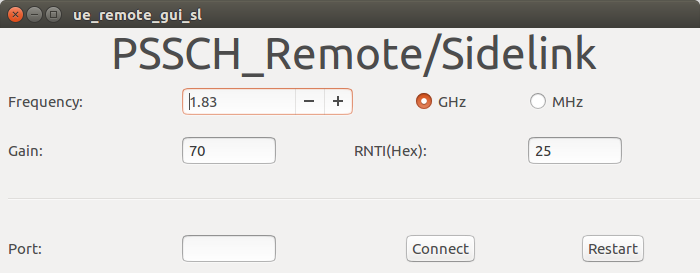}  
	\caption{Controller GUI for the remote UE using the sidelink.}
	\label{fig:remote_sidelink_GUI}
\end{figure}

The controller \ac{GUI} for the remote \ac{UE} comprises three \ac{GUI} windows. The first \ac{GUI} acts as the controller through which the mode selection between downlink and sidelink can be performed (shown in Fig.~\ref{fig:remote_controller}). 
If the experimenter selects the downlink mode, the \ac{LTE} radio software for the downlink automatically boots up and the \ac{GUI} for the downlink opens up (as shown in Fig.~\ref{fig:remote_downlink_GUI}), and similarly for the sidelink mode (as shown in Fig.~\ref{fig:remote_sidelink_GUI}. The parameters which the experimenter can control through downlink or sidelink \ac{GUI} are similar to those of the downlink GUI for the relay \ac{UE}. 

\section{Conclusion}
\label{sec:conclusion}

In this paper, we have implemented the first \ac{LTE} standard-compliant \ac{SDR} sidelink \ac{PHY} layer data transmission. We have built a relay \ac{UE} to relay the data received from an eNodeB (through the downlink) to a remote \ac{UE} (through the sidelink). We have also developed functionality for mode selection between infrastructure mode and \ac{D2D} mode on the remote \ac{UE}, providing experimenters with a \ac{GUI} that allows them to perform mode selection and change the radio parameters on the fly. Our measurement results show that the sidelink relay can provide range extension for the cell.

Our ongoing work includes the development of  intelligent mode selection (UE-to-UE versus UE-to-eNodeB) according to network conditions, as well as enhanced \ac{PSCCH} functionality, including user discovery and resource management.

\begin{acronym} 

\acro{3GPP}{Third Generation Partnership Project}
\acro{BLER}{Block Error Rate}
\acro{D2D}{Device-to-Device}
\acro{DFT}{Discrete Fourier Transform}
\acro{DSP}{Digital Signal Processing} 
\acro{eWINE}{Elastic Wireless Network Experimentation}
\acro{FDD}{Frequency Division Duplex}
\acro{GUI}{Graphical User Interface}
\acro{IDFT}{Inverse Discrete Fourier Transform}
\acro{LTE}{Long Term Evolution}
\acro{MAC}{Medium Access Control}
\acro{MCS}{Modulation and Coding Scheme}
\acro{MIB}{Master Information Block}
\acro{OFDM}{Orthogonal Frequency Division Multiplexing}
\acro{PDCCH}{Physical Downlink Control Channel}
\acro{PDSCH}{Physical Downlink Shared Channel}
\acro{PHY}{Physical}
\acro{PRBs}{Physical Resource Blocks}
\acro{PSBCH}{Physical Sidelink Broadcast Channel}
\acro{PSCCH}{Physical Sidelink Control Channel}
\acro{PSDCH}{Physical Sidelink Discovery Channel}
\acro{PSS}{Primary Synchronization Signal}
\acro{PSSCH}{Physical Sidelink Shared Channel}
\acro{PSSS}{Primary Sidelink Synchronization Signal}
\acro{QAM}{Quadrature Amplitude Modulation}
\acro{QoS}{quality of service}
\acro{RNTI}{Radio Network Temporary Identifier}
\acro{RSSI}{received signal strength indicator}
\acro{SC-FDMA}{Single Carrier Frequency Division Multiple Access}
\acro{SCI}{Sidelink Control Information}
\acro{SDMA}{Space Division Multiple Access}
\acro{SDN}{Software Defined Networks}
\acro{SDR}{Software-Defined Radio}
\acro{SNR}{Signal-to-Noise Ratio}
\acro{TDMA}{Time Division Multiple Access}
\acro{UE}{User Equipment}
\acro{UHD}{USRP Hardware Driver}
\acro{USRP}{Universal Software Radio Peripheral}

\end{acronym}

\section*{Acknowledgment}
This publication has emanated from research supported by the European Commission Horizon 2020 Programme under grant agreement no. 688116 (eWINE). We also thank Mr. Robson Couto for his effort on developing the first version of Graphical User Interface, and Software Radio System (SRS) for technical support on srsLTE software.

\bibliographystyle{IEEEtran}
\bibliography{bibliography}
\end{document}